\documentclass[twocolumn,aps,preprintnumbers,amsmath,amssymb,superscriptaddress,nobibnotes]{revtex4}
\usepackage{graphicx}
\usepackage{dcolumn}
\usepackage{bm}
\usepackage{textcomp}
\usepackage{units}
\usepackage{graphics}
\usepackage{longtable}

\begin{document}
\setlength{\parindent}{0em} 

\title{Electron matter wave interferences at high vacuum pressures}

\author{G. Sch\"{u}tz$^1$, A. Rembold$^1$, A. Pooch$^1$, W.T. Chang$^2$ and A. Stibor}
\email{alexander.stibor@uni-tuebingen.de}
\affiliation{Institute of Physics and Center for Collective Quantum Phenomena in LISA$^+$,
University of T\"{u}bingen, Auf der Morgenstelle 15, 72076 T\"{u}bingen, Germany\\
$^2$Institute of Physics, Academia Sinica, Nankang, Taipei, Taiwan, Republic of China}

\begin{abstract}

The ability to trap and guide coherent electrons is gaining importance in fundamental as well as in applied physics. In this regard novel quantum devices are currently developed that may operate under low vacuum conditions. Here we study the loss of electron coherence with increasing background gas pressure. Thereby, optionally helium, hydrogen or nitrogen is introduced in a biprism interferometer where the interference contrast is a measure for the coherence of the electrons. The results indicate a constant contrast that is not decreasing in the examined pressure range between \unit[$10^{-9}$]{mbar} and \unit[$10^{-4}$]{mbar}. Therefore, no decoherence was observed even under poor vacuum conditions. Due to scattering of the electron beam with background H$_2$-molecules a signal loss of \unit[94]{\%} was determined. The results may lower the vacuum requirements for novel quantum devices with free coherent electrons.

\end{abstract} 

\maketitle

\section{Introduction}

\begin{figure*}[t]
\centerline{\scalebox{0.41}{\includegraphics{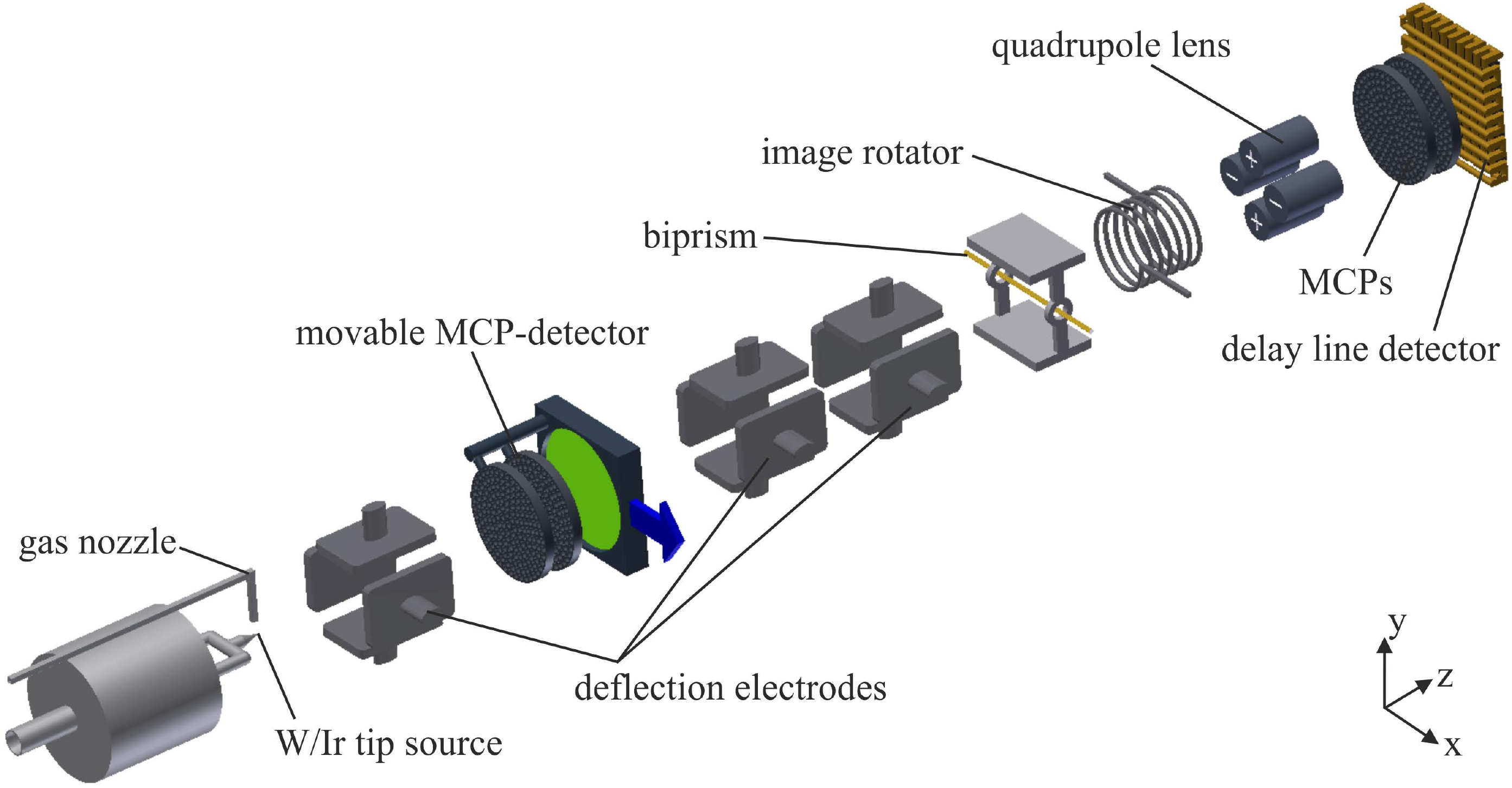}}}
\caption{(Color online) Experimental setup of the biprism interferometer to test the electron coherence for increasing background gas pressure (not to scale) \cite{Schuetz2014,Hasselbach1998a,Hasselbach2010a,Maier1997}. }
\label{figure1}
\end{figure*}

The coherent control and interference of free electrons has a long history. In the 1950s a mayor scientific breakthrough happened with the development of biprism electron interferometers \cite{Moellenstedt1955}. A variety of experiments for free electrons were accomplished in the following decades proving e.g.~the magnetic Aharonov-Bohm effect \cite{Moellenstedt1962}, the Sagnac effect \cite{Hasselbach1993} or Hanbury Brown-Twiss anticorrelations \cite{Kiesel2002}. In recent years the coherent control of free electrons is gaining again importance for fundamental research and in a technical point of view. This can be observed in decoherence studies of electrons near semiconducting surfaces \cite{Sonnentag2007} and developments such as a field emission source for free electron femtosecond pulses \cite{Hommelhoff2006a,Hommelhoff2006b}, surface-electrode chips \cite{Hammer2014} or a biprism electron interferometer with a single atom tip source \cite{Schuetz2014}. New quantum devices with coherent electrons are currently implemented like a recently proposed noninvasive quantum electron microscope \cite{Putnam2009}. Due to the quantum Zeno effect it potentially reduces the electron radiation exposure during scanning of fragile biological samples by two orders of magnitude.  \\ 
\\
Some of these applications may operate under low vacuum conditions or are technically less demanding to realize if an ultra-high vacuum (UHV) environment is not needed. The question arises, how background gases influence the properties of the matter wave. Important applications such as reflection high energy electron diffraction (RHEED) for in situ monitoring of the growth of thin films on surfaces \cite{Rijnders1997} or ultrafast electron diffraction (UED) at molecular beams for direct imaging of transient molecular structures \cite{Goodson2003,Ihee2001} are known to work at high background gas pressures. \\
However, it has not been studied yet, how the coherence of an electron beam is influenced by increasing background gas pressure. The gradual loss of coherence through collisions with background gases was analyzed for neutral C$_{60}$-fullerenes in a near field Talbot-Lau matter wave interferometer \cite{Hornberger2003}. Thereby, decoherence was observed at a gas pressure of $\sim$ \unit[$10^{-6}$]{mbar}.\\
\\
In this work we study the possible loss of coherence for electron matter waves in a biprism interferometer in presence of helium (He), nitrogen (N$_2$) or hydrogen (H$_2$) background gas. Our instrument is able to generate interferograms with high interference contrast in a pressure range between $10^{-9}$ and \unit[$10^{-4}$]{mbar}. It is only limited by the vacuum specifications of the multi channel plate (MCP) detector. We will demonstrate that in this whole pressure region no decoherence can be observed.

\section{Setup}
\label{setup}

A scheme of our experimental setup is shown in Fig.~\ref{figure1} and is described in detail elsewhere \cite{Schuetz2014,Hasselbach1998a,Hasselbach2010a,Maier1997}. In our approach the electron beam is field emitted by an etched tungsten tip that is covered with a monolayer of iridium and annealed to form a protrusion in the nanometer regime \cite{Kuo2006a,Kuo2008}. The tip forming procedure is monitored by a MCP-detector that can be moved out of the optical axis. The electrons start with an emission energy of \unit[1.58]{keV} for the experiment with He and N$_2$ background gas and \unit[1.44]{keV} for the one with H$_2$. They coherently illuminate a \unit[400]{nm} wide gold covered biprism fiber that divides and combines the electron matter waves \cite{Moellenstedt1955,Schuetz2014}. It is set on a positive potential of \unit[0.35]{V} for the experiments with He or N$_2$ background gas and \unit[0.4]{V} for H$_2$. All beam alignment is performed by electrostatic deflection electrodes. Behind the biprism the partial waves overlap and interfere with each other. The interference pattern has a period of several \unit[100]{nm} and is oriented parallel to the biprism in the $x$-direction. It is magnified by an electrostatic quadrupole lens to fit the detectors resolution of about \unit[100]{\textmu m}. The image rotator, a magnetic coil, allows to rotate the interferogram to correct possible misalignments. The interference pattern is detected by a MCP-detector with a delay line anode. It is able to operate at background gas pressures up to about \unit[$10^{-4}$]{mbar}. Above that level the risk of destruction of the MCPs due to electric discharges is high. 
\\
\begin{figure}[t]
\centerline{\scalebox{1}{\includegraphics{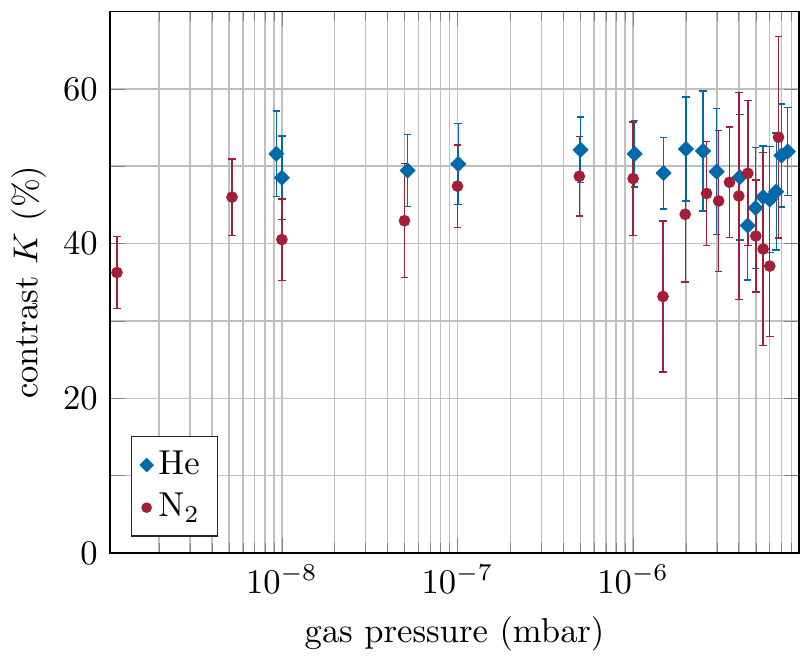}}}
\caption{(Color online) Electron interference contrast $K$ as a function of increasing pressure of helium (He, blue squares) or nitrogen (N$_2$, red dots) background gas.}
\label{figure2}
\end{figure}

The whole interferometer has a length (tip to detector) of \unit[565]{mm}. It is constructed rigidly \cite{Hasselbach1988} to avoid mechanical vibrations and shielded against electromagnetic noise \cite{Rembold2014} by a copper and mu-metal tube. The inlet of different background gases is performed by an UHV gas nozzle. The interferometer is placed in a chamber where a minimum pressure of \unit[$1 \times 10^{-10}$]{mbar} is achieved by an ion getter pump in combination with a cryopump.

\section{Measurements}

Three experimental runs were performed introducing either He, N$_2$ or H$_2$ gas in the UHV chamber. Before each run, the tip was annealed to form a protrusion being the emission center. Therefore, possible variations in the tip apex size influencing the electron emission voltage and the maximal contrast cannot be ruled out. The measurement started at a background gas pressure of \unit[$10^{-9}$]{mbar} and electron interferograms were recorded with a signal acquisition time of \unit[25]{s} for He or N$_2$ and \unit[22]{s} for H$_2$. Then a further small amount of gas was introduced through the nozzle. At equilibrium another interference pattern was recorded. This process was repeated stepwise with increasing pressure for the different gases. Background images for the same integration time were acquired in the experiments with He or N$_2$ by switching off the field emission subsequently to each recording. This data was subtracted from the interferograms. For the H$_2$ measurement no background subtraction was necessary since the ion getter pump, as the main source of background, was turned off.
The recorded images were analyzed by adding all counts in the pixel-rows of the detector along the $x$-direction of the interference pattern and dividing the sum by the amount of pixel-columns. The distribution of the resulting average interference pattern $I(y)$ versus the $y$-direction normal to the interference stripes was fitted by the following expression to determine the mean intensity $I_0$, the pattern periodicity $d_s$ and the contrast $K$ \cite{Lenz1984}

\begin{equation}
I(y) = I_0 \left[1 + K \cos\left(\frac{2 \pi y}{d_s}\right) \right].
\label{eq:fit}
\end{equation}

\begin{figure}[t]
\centerline{\scalebox{1}{\includegraphics{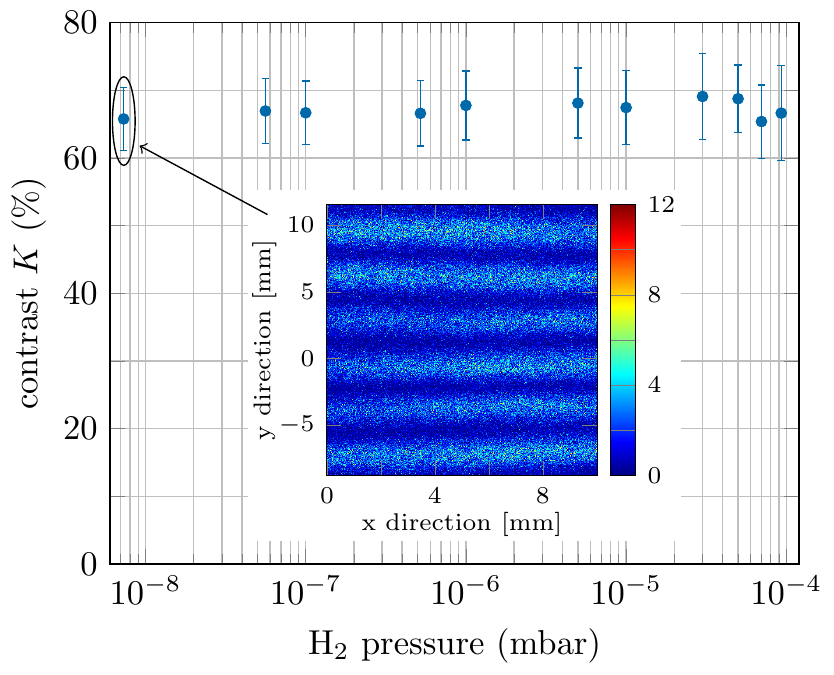}}}
\caption{(Color online) Electron interference contrast $K$ as a function of increasing pressure of hydrogen (H$_2$) background gas. Inset: Electron interference pattern at a H$_2$ gas pressure of \unit[$7.3~\times~10^{-9}$]{mbar}.}
\label{figure3}
\end{figure}

In Fig.~\ref{figure2} the resulting contrast is plotted versus the background gas pressure for He and N$_2$. The contrast distribution is rather constant for the whole measured pressure range indicating the electrons remain coherent. At higher pressures the ion pump produced an increasing noise level of ions on the detector that leads to greater error bars and a higher dispersion of the data points. For hydrogen we were able to work without the ion getter pump and stabilized the pressure only with the cryopump. This significantly reduced the background counts resulting in a more stable signal. The pressure-dependent contrast for hydrogen is shown in Fig.~\ref{figure3}. It is constant around \unit[67]{\%}. The inset illustrates a typical interference pattern recorded at \unit[$7.3~\times~10^{-9}$]{mbar}.\\

\begin{figure}[t]
\centerline{\scalebox{1}{\includegraphics{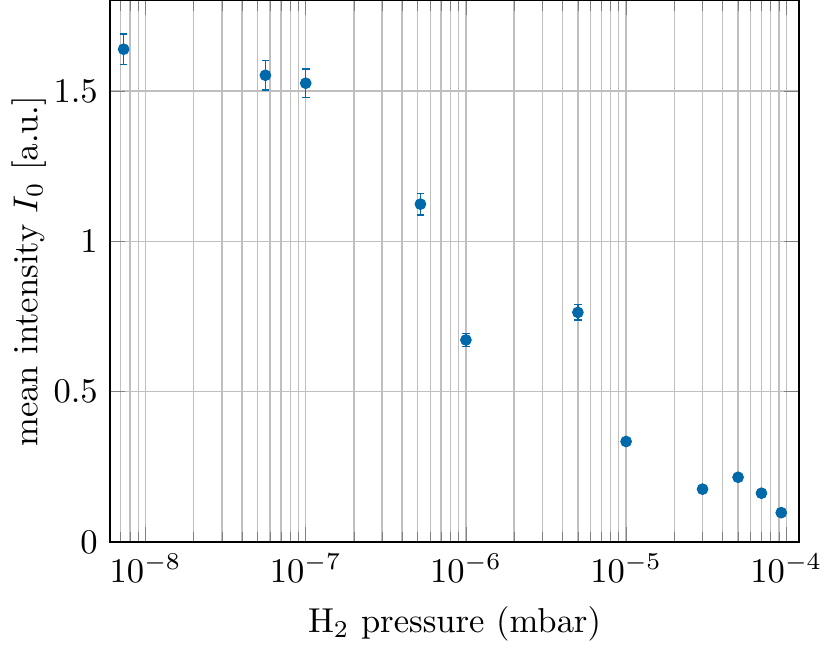}}}
\caption{Mean electron intensity on the detector as a function of increasing pressure of H$_2$ background gas.}
\label{figure4}
\end{figure}

Additionally, the mean intensity $I_0$ of the interference pattern on the detector was determined for hydrogen with increasing pressure. It represents the center line of the cosine-function in Eq.~\ref{eq:fit}. The data is plotted in Fig.~\ref{figure4}. As expected, a significant signal drop is determined. This is presumably due to increasing collisions between electrons and H$_2$ molecules that decrease the count rate. At a pressure of \unit[$9.3 \times 10^{-5}$]{mbar} only a fraction of \unit[6]{\%} of the original electron signal is left.\\

\section{Discussion and Conclusion}
\label{conclusion}

We have studied the coherent properties of electron matter waves in a biprism interferometer under low vacuum conditions by introducing helium, nitrogen or hydrogen background gas in the UHV chamber. Unlike to interference experiments with C$_{60}$ fullerenes \cite{Hornberger2003} the electrons in our instrument do not show decoherence up to a pressure of \unit[$\sim 10^{-4}$]{mbar} which can be observed in a constant interference contrast. In the C$_{60}$ near field interferometer the heavy molecules have a significant probability to be measured in the region between the interference stripes after a collision, leading to a loss of contrast. In our far field interferometer the situation is different. After a collision with a significantly heavier background gas atom or molecule, the electron is in most cases scattered into an angle large enough to miss the detector. Due to the quadrupole magnification, minimal deflection of the electrons trajectory lead to a large displacement in the detection plane. This is indicated by the significant signal loss of \unit[94]{\%} comparing the mean electron intensity measured on the detector at a H$_2$ pressure of \unit[$7.3 \times 10^{-9}$]{mbar} with the one at \unit[$9.3 \times 10^{-5}$]{mbar}. In other words, those electrons that made it to the detector did not scatter on a gas atom and are therefore still coherent. It is an advantage of this setup to be able to select these electrons and remain high contrast interference pattern even under rather low vacuum conditions. \\
\\
Due to possible electric discharges in the MCP-detector, interference at even higher vacuum pressures could not be studied. However, coherent behaviour of electrons at a comparable pressure of \unit[$\sim \,3 \times 10^{-4}$]{mbar} was reported in an UED experiment \cite{Goodson2003} and for significantly higher pressures in a RHEED measurement \cite{Rijnders1997}. The latter describes electron diffraction on SrTiO$_3$ and YBa$_2$Cu$_3$O$_{7\,-\,\delta}$ surfaces at an oxygen background pressure up to \unit[0.15]{mbar}. This was possible due to a short design of the setup, a differential pumping unit for the source and significantly higher electron energies of \unit[35]{keV} that allowed to work without MCP amplification prior to the detection on the fluorescent screen. In accordance to our observations, also a strong scattering loss of electrons in the high oxygen pressure was observed in the RHEED experiment \cite{Rijnders1997}. \\
We therefore conclude that with different detection methods and shorter configurations, electron diffraction or interference may be observed at even higher background gas pressures. The results of our experiments provide an indication of the vacuum requirements for novel devices applying free coherent electrons. 

\section{Acknowledgements}

This work was supported by the Deutsche Forschungsgemeinschaft (DFG, German Research Foundation) through the Emmy Noether program \mbox{STI 615/1-1.} A.R. acknowledges support from the Evangelisches Studienwerk e.V. Villigst.

\end{document}